\documentclass[usenatbib]{mnras}
\usepackage{graphicx}

\title[Distinguishing Between Stellar and Planetary Companions]
  {Distinguishing Between Stellar and Planetary Companions With
  Phase Monitoring}
\author[Stephen R. Kane \& Dawn M. Gelino]
  {Stephen R. Kane, Dawn M. Gelino\\
  NASA Exoplanet Science Institute, Caltech, MS 100-22, 770 South
  Wilson Avenue, Pasadena, CA 91125, USA}

\begin{document}

\maketitle

%%%%%%%%%%%%%%%%%%%%%%%%%%%%%%%%%%%%%%%%%%%%%%%%%%%%%%%%%%%%%%%%%%%%

\begin{abstract}

Exoplanets which are detected using the radial velocity technique have
a well-known ambiguity of their true mass, caused by the unknown
inclination of the planetary orbit with respect to the plane of the
sky. Constraints on the inclination are aided by astrometric follow-up
in rare cases or, in ideal situations, through subsequent detection of
a planetary transit. As the predicted inclination decreases, the mass
of the companion increases leading to a change in the predicted
properties. Here we investigate the changes in the mass, radius, and
atmospheric properties as the inclination pushes the companion from
the planetary into the brown dwarf and finally low-mass star
regimes. We determine the resulting detectable photometric signatures
in the predicted phase variation as the companion changes properties
and becomes self-luminous. We apply this to the HD~114762 and
HD~162020 systems for which the minimum masses of the known companions
places them at the deuterium-burning limit.

\end{abstract}


%%%%%%%%%%%%%%%%%%%%%%%%%%%%%%%%%%%%%%%%%%%%%%%%%%%%%%%%%%%%%%%%%%%%

\section{Introduction}
\label{intro}

Planets discovered using the radial velocity technique continue to
form a major component of the known exoplanets. These planets have a
well-known ambiguity to their masses due to the unknown inclination of
their orbits. Some planets have subsequently been found to be more
massive than originally thought when constraints are later placed upon
their inclinations. These constraints can come from dynamical
considerations, such as for the HD~10180 system \citep{lov11}, through
astrometric follow-up \citep{ref11}, or measurements of the projected
equatorial velocity \citep{wat10}, although some hot Jupiters have
been found to exhibit spin-orbit misalignment (see for example
\citet{win05}). The consequence of these constraints can be either to
confirm their planetary candidacy or to move the mass into the regime
of brown dwarfs and low-mass stars.

The mass-radius relationship of short-period exoplanets is evolving
through the discovery of numerous transiting exoplanets
\citep{bur07,for07,sea07}. The understanding of this relationship is
undergoing continued and rapid evolution through the release of
transiting planets from the Kepler mission
\citep{bor11a,bor11b}. Surveys for transiting exoplanets have also led
to the serendipitous discovery of transiting brown dwarfs, such as
CoRoT-3b \citep{del08}, WASP-30b \citep{and11}, and LHS6343C
\citep{joh11}. The mass-radius relationship of low-mass stars has been
investigated by numerous authors \citep{dem09,fer09,kra11,rib06}
through new radii measurements and the development of models to
explain the observed relationship. Even so, the low-mass stars for
which accurate radii have been determined remains a relatively small
sample from which to draw upon to derive the theoretical framework for
model construction. These mass-radii distributions give us clues about
the formation mechanisms that occurred within these systems.

The phase functions and resulting photometric light curves of orbiting
companions depends upon their radii and albedo as well as the orbital
components \citep{kan10,sea00,sud05}. \citet{kan11a} showed how the
phase curves of exoplanets varies with inclination. However, it was
assumed that the fundamental physical properties of the planet (such
as mass, radius, and atmospheric properties) remain the same. This is
particularly relevant to planets discovered using the radial velocity
technique since these planets are inherently subjected to an ambiguity
in the mass with respect to the unknown orbital inclination. Here we
expand upon this topic to investigate how the properties of a known
companion can be expected to change as the inclination is decreased
and therefore the mass is increased. If the companions are
self-luminous then this places an extra constraint upon what one can
expect to see in high-precision observations designed to measure phase
variations. We thus produce a criteria from which one can distinguish
between stellar and planetary companions solely from high-precision
photometric monitoring and without the need for astrometric
observations.

%%%%%%%%%%%%%%%%%%%%%%%%%%%%%%%%%%%%%%%%%%%%%%%%%%%%%%%%%%%%%%%%%%%%

\section{Increasing the Companion Mass}

In this section, we describe the changing properties of a companion as
the mass increases from the planetary regime, through the brown dwarf
regime, and into the realm of low-mass stars. We further investigate
how these properties influence the photometric properties up to the
self-luminous threshold.

%%%%%%%%%%%%%%%%%%%%%%%%%%%%%%%%%%%%%%%%%%%%%%%%%%%%%%%%%%%%%%%%%%%%

\subsection{Mass-Radius Dependence}
\label{massradius}

The mass-radius relationship of exoplanets, brown dwarfs, and low-mass
stars is a difficult subject to relate to measurements since these
require a system where the orbital inclination allows for the
observations of eclipses and transits. Locating such systems involves
the monitoring of many stars since the probability of having a
favourable orbital inclination tends to be relatively low, depending
upon the semi-major axis of the orbiting companion.

The radii for objects of a given mass depends upon a number of
factors, including age and metallicity. Here we consider host stars
which comprise the bulk of the known radial velocity hosts which have
typical ages in the range 1--5~Gyrs \citep{tak07}. A study of 49
exoplanet host stars by \citet{saf05} found median ages of 5.2 and
7.4~Gyrs, using chromospheric and isochrone methods respectively, with
dispersions of $\sim 4$~Gyrs. We thus consider objects which are of
order 5~Gyrs old.

The true mass of companions detected using the radial velocity
technique is the measured mass divided by $\sin i$ where $i$ is the
orbital inclination relative to the plane of the sky. As the
inclination decreases from an edge-on configuration ($i = 90\degr$),
the true mass increases which in turn effects the radius and
atmospheric properties of the companion. It is generally held that
the separation between planets and brown dwarfs is the onset of
deuterium burning. However, the mass criteria for determining where
deuterium burning occurs can be quite broad depending upon
helium/deuterium abundances and metallicity \citep{spi11}.

\begin{figure}
  \includegraphics[angle=270,width=8.2cm]{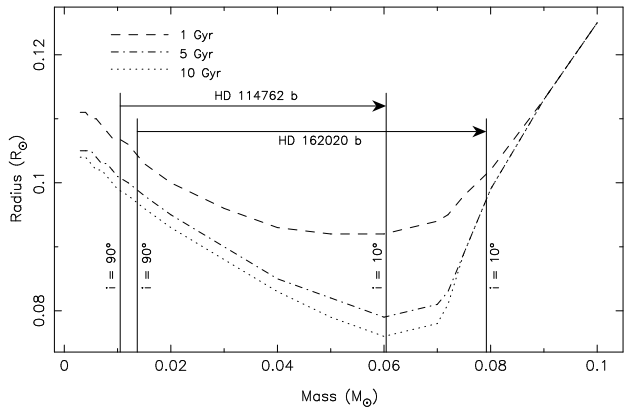}
  \caption{The mass-radius relationship of planets through to low-mass
    stars using the isochrones of \citet{bar03} for 1, 5, and 10
    Gyrs. The increasing masses of HD~114762b and HD~162020b are shown
    to demonstrate the effect of decreasing the inclination from an
    edge-on orbit ($i = 90\degr$) to an almost face-on orbit ($i =
    10\degr$).}
  \label{isochrones}
\end{figure}

By way of demonstration, Figure \ref{isochrones} plots the COND
evolutionary model isochrones of \citet{bar03} for 1, 5, and 10 Gyrs
and for a mass range of 0.003--0.100~$M_\odot$. Also shown are the
masses of two known exoplanets, HD~114762b \citep{lat89} and
HD~162020b \citep{udr02}. Assuming an inclination of $i = 90\degr$,
the measured mass of HD~114762b is 11.0~$M_J$ (0.0105~$M_\odot$)
whereas for a near face-on inclination of $i = 10\degr$ the mass
increases to 63.2~$M_J$ (0.0604~$M_\odot$). According to the 5~Gyr
isochrone, this reduces the predicted radius from 0.101~$R_\odot$ to
0.079~$R_\odot$. \citet{kan11c} have excluded transits for this
companion which thereby restricts the inclination to be less than
$89\degr$. For HD~162020b, the same change in inclination increases
the measured mass of 14.4~$M_J$ (0.0137~$M_\odot$) to 82.9~$M_J$
(0.0792~$M_\odot$), thus decreasing the predicted radius from
0.099~$R_\odot$ to 0.097~$R_\odot$. In each case, the change in
inclination moves them from the planetary regime to the brown
dwarf/low-mass star boundary and beyond.

As noted in Section \ref{intro}, our knowledge of companion radii in
different mass regimes is a currently evolving topic. For example, an
important consideration is the existence of the so-called brown dwarf
desert \citep{gre06,kra08} which may bias inclinations away from this
mass regime. The COND evolutionary models utilized above will
undoubtedly undergo slight adjustments as the sample size of known
low-mass transiting companions increases. One of the important
parameters is the age of the host star. As shown in Figure
\ref{isochrones}, the range of radii can be quite diverse for a given
mass depending on which isochrone one adopts for the host
star. However, the divergence is most significant for relatively young
($\sim 1$~Gyr) stars beyond which the radii converge upon a small
range of radii for a large range of ages. Even so, one must consider
the uncertainty in the host star age, assuming that the companion is
of a similar age.

%%%%%%%%%%%%%%%%%%%%%%%%%%%%%%%%%%%%%%%%%%%%%%%%%%%%%%%%%%%%%%%%%%%%

\subsection{Impact on Phase Curves}

For companions which are not self-luminous, there will be a
photometric phase signature from the companion whose amplitude depends
upon a variety of factors including the companions radius, semi-major
axis, eccentricity, orbital inclination, and atmospheric properties
(geometric albedo). Here we adopt the formalism of \citet{kan10} and
\citet{kan11a} to demonstrate the impact of the changing companion
properties on phase curves.

The phase angle of the planet ($\alpha$) is given by
\begin{equation}
  \cos \alpha = \sin (\omega + f) \sin i
  \label{phaseangle}
\end{equation}
where $\omega$ is the argument of periastron and $f$ is the true
anomaly. This angle is defined to be $\alpha = 0\degr$ when the planet
is on the opposite side of the star from the observer. Note that this
means that $\cos \alpha = 1$ is only possible for edge-on orbits. All
other inclinations will result in a more complicated dependence on the
orbital parameters as described by \citet{kan11a}.

The flux of a planet $f_p$ and host star $f_\star$ has a ratio defined
as
\begin{equation}
  \epsilon(\alpha,\lambda) \equiv
  \frac{f_p(\alpha,\lambda)}{f_\star(\lambda)}
  = A_g(\lambda) g(\alpha,\lambda) \frac{R_p^2}{r^2}
  \label{fluxratio}
\end{equation}
where the flux is measured at wavelength $\lambda$, $A_g(\lambda)$ is
the geometric albedo, $g(\alpha,\lambda)$ is the phase function, and
$R_p$ is the radius of the planet. The star--planet separation $r$ is
given by
\begin{equation}
  r = \frac{a (1 - e^2)}{1 + e \cos f}
  \label{separation}
\end{equation}
where $a$ is the semi-major axis and $e$ is the orbital
eccentricity. The phase function is primarily dependent upon the phase
angle, whether one assumes a Lambert sphere or something more
complicated to describe the scattering properties. The geometric
albedo for gas giant planets depends upon the incident flux
(star--planet separation) since this can determine the amount of
reflective condensates that can maintain a presence in the upper
atmospheres\citep{kan10,sud05}.

A component of Equation \ref{fluxratio} which explicitly relies upon
the companion properties is the radius squared. As shown in Section
\ref{massradius}, variation of the companion mass in the range
0.003--0.100~$M_\odot$ can vary the radius by as much as $\sim
20$\%. This results in a variation of the flux ratio amplitude by as
much as $\sim 40$\%. The uncertainty in both the planetary radius and
albedo are discussed further in Section \ref{unknown}. For now we note
that there is an apparent degeneracy in the predicted radius as a
function of mass in the isochrones shown in Figure \ref{isochrones}
due to the minimum located at 0.06~$R_\odot$. However, beyond this
value the companions start to become self-luminous. This is discussed
in further detail in Section \ref{luminous}.

%%%%%%%%%%%%%%%%%%%%%%%%%%%%%%%%%%%%%%%%%%%%%%%%%%%%%%%%%%%%%%%%%%%%

\subsection{Ellipsoidal Variations}
\label{ellipvar}

An additional possible effect of an increased mass is the induction of
ellipsoidal variations in the host star. \citet{dra03} have proposed
the use of ellipsoidal variations as a tool in eliminating false
positives due to eclipsing binaries from transit surveys. Stellar
ellipsoidal variations have been previously detected for short-period
giant planet systems though, such as the case of HAT-P-7b which was
observed to have signatures of phase as well as ellipsoidal variations
in Kepler data \citet{wel10}. The detection of ellipsoidal variations
in the context of the Kepler mission is discussed in detail by
\citet{pfa08}.

\begin{figure}
  \includegraphics[angle=270,width=8.2cm]{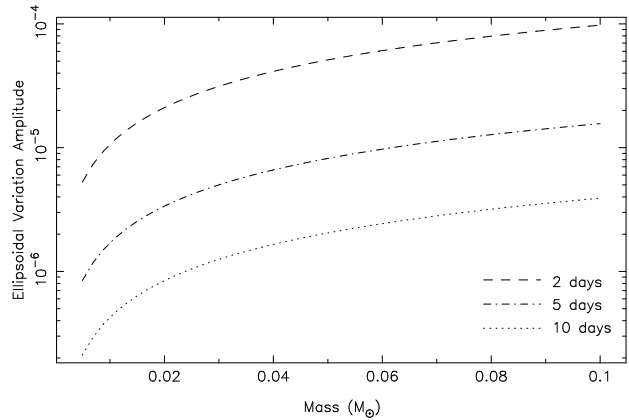}
  \caption{The mass dependence of the predicted ellipsoidal variation
    amplitude and the same range of masses as shown in Figure
    \ref{isochrones} for three different periods.}
  \label{ellip}
\end{figure}

Comparison of phase and ellipsoidal variation amplitudes to those
induced by reflex Doppler motion has been undertaken in several
studies, including those by \citet{loe03} and \citet{zuc07}. A full
expansion of the discrete Fourier series for the ellipsoidal variation
amplitude is provided by \citet{mor93}. Here we adopt the approximate
relation used by \citet{loe03} as follows
\begin{equation}
  \frac{\Delta F}{F_0} \sim \beta \frac{M_p}{M_\star} \left(
  \frac{R_\star}{a} \right)^3
  \label{ellipsoidal}
\end{equation}
where $\beta$ is the gravity darkening exponent. For the purposes of
demonstration, we have assumed $\beta = 0.32$ as determined by
\citet{luc67}. We refer the reader to \citet{cla00} for a much more
thorough treatment of the dependence of $\beta$ on various stellar
properties. Figure \ref{ellip} shows how the ellipsoidal variation
amplitude can be expected to change as a function of companion mass
for three different orbital periods. The scale used on the x-axis is
identical to that in Figure \ref{isochrones}. As one would expect from
Equation \ref{ellipsoidal}, a change in the companion mass results in
an almost linear change in the observed flux ratio changes (``almost''
since changing the mass also changes the semi-major axis at which the
expected period occurs). Changing the period has a much more dramatic
effect. Note however that a low-mass star in a 10 day orbit will
induce a similar ellipsoidal variation to a giant planet in a 5 day
orbit.

There are numerous examples of a changing tidal distortion amplitude
due to eccentric orbits, such as those contained in \citet{sos04}.  To
account for orbital eccentricity and inclination, we modify Equation
\ref{ellipsoidal} as follows:
\begin{equation}
  \frac{\Delta F}{F_0} \sim \beta \frac{M_p}{M_\star} \left(
  \frac{R_\star}{r} \right)^3 \left( \cos^2 (\omega + f) + \sin^2
  (\omega + f) \cos^2 i \right)^{\frac{1}{2}}
  \label{ellipecc}
\end{equation}
where we have replaced $a$ with $r$, as per Equation \ref{separation}.
If there is significant orbital eccentricity present, one can use this
as an additional diagnostic in interpreting the uniqueness of the
ellipsoidal variation solution. However, eccentric orbits are far more
likely to occur at longer periods where the amplitude of the variation
is greatly reduced.

%%%%%%%%%%%%%%%%%%%%%%%%%%%%%%%%%%%%%%%%%%%%%%%%%%%%%%%%%%%%%%%%%%%%

\section{Crossing the Self-Luminous Threshold}
\label{luminous}

Moving from the brown dwarf into the low-mass star regime introduces
many changes associated with the companion becoming self-luminous.
Planets and brown dwarfs are typically not self-luminous except at a
young age ($< 1$~Gyr). For brown dwarfs engaged in deuterium burning,
the transition to hydrogen burning will abort if the mass is less than
$\sim$0.07~$M_\odot$ after which the object will undergo luminosity
decay beyond an age of $\sim$1~Gyr. A full description of object
characteristics in terms of age, helium/deuterium fraction, and
metallicity can be found in \citet{bur93}. Here we briefly discuss the
main features that one can expect to observe as the object crosses the
self-luminous threshold.

%%%%%%%%%%%%%%%%%%%%%%%%%%%%%%%%%%%%%%%%%%%%%%%%%%%%%%%%%%%%%%%%%%%%

\subsection{Elimination of the Phase Function}

A major feature of planetary phase functions is that it relies on
reflected light which is incident from the host star. If the dominant
source of light from the companion becomes that which originates from
self-luminosity then there will no longer be a phase function of the
kind described earlier. To account for this, we include a gradual
decrease in the phase variation amplitude for companion masses $>
50$~$M_J$. There will however be an increasing likelihood of
observable ellipsoidal variations with increasing mass (see Section
\ref{ellipvar}) such that these variations may subsume the decreasing
phase effects.

An additional effect which may produce a phase-like signature is the
``reflection effect''; a well-known phenomenon for close binary stars
in which the irradiation of one component by the other results in a
differential flux over the irradiated stars surface
\citep{wil90}. Application of this process to binary systems
consisting of a star and a giant planet or brown dwarf is considered
by \citet{bud11}, although \citet{gre03} argue that this heating of
the primary star due to the presence of even a close-in gas giant
planet will be negligibly small. The irradiation will only become
comparable to the amplitude of the phase variation when the mass
enters the low-mass star regime, at which point the convective
atmosphere will in turn have an increased albedo due to irradiation
from the primary \citep{har03}.

For planetary and brown dwarf companions, the heating of the day-side
may result in significant thermal flux being re-radiated. This depends
not only on the incident flux but also the fundamental atmospheric
properties which govern radiative and advective time scales
\citep{for08,kan11b,sho09}. Thus efficient atmospheric re-circulation
of the received heat to the night-side of the planet may be sufficient
to suppress an overall phase function at infra-red wavelengths.

%%%%%%%%%%%%%%%%%%%%%%%%%%%%%%%%%%%%%%%%%%%%%%%%%%%%%%%%%%%%%%%%%%%%

\subsection{Introduction of Stellar Activity}

\begin{figure*}
  \begin{center}
    \begin{tabular}{cc}
      \includegraphics[angle=270,width=8.2cm]{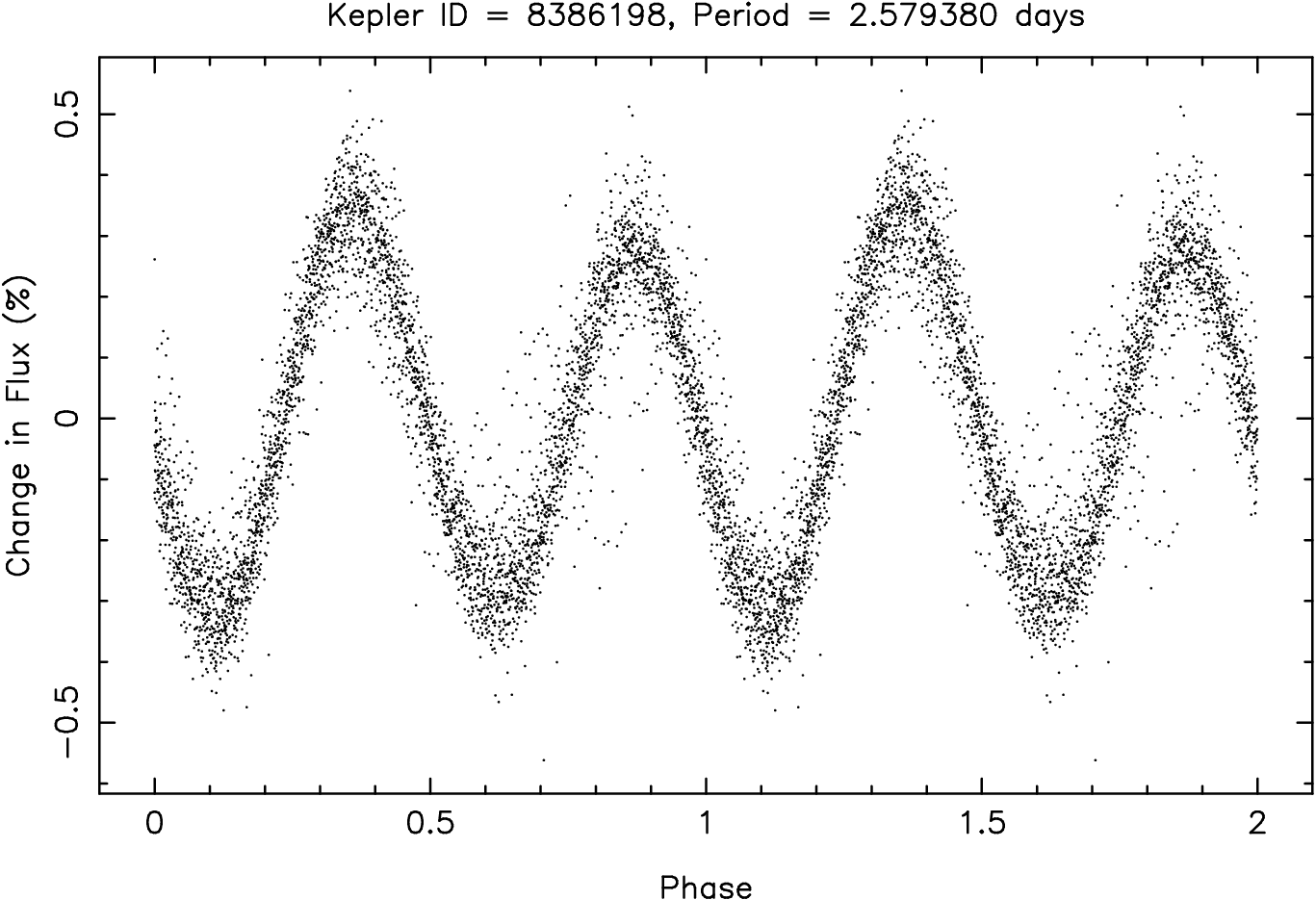} &
      \includegraphics[angle=270,width=8.2cm]{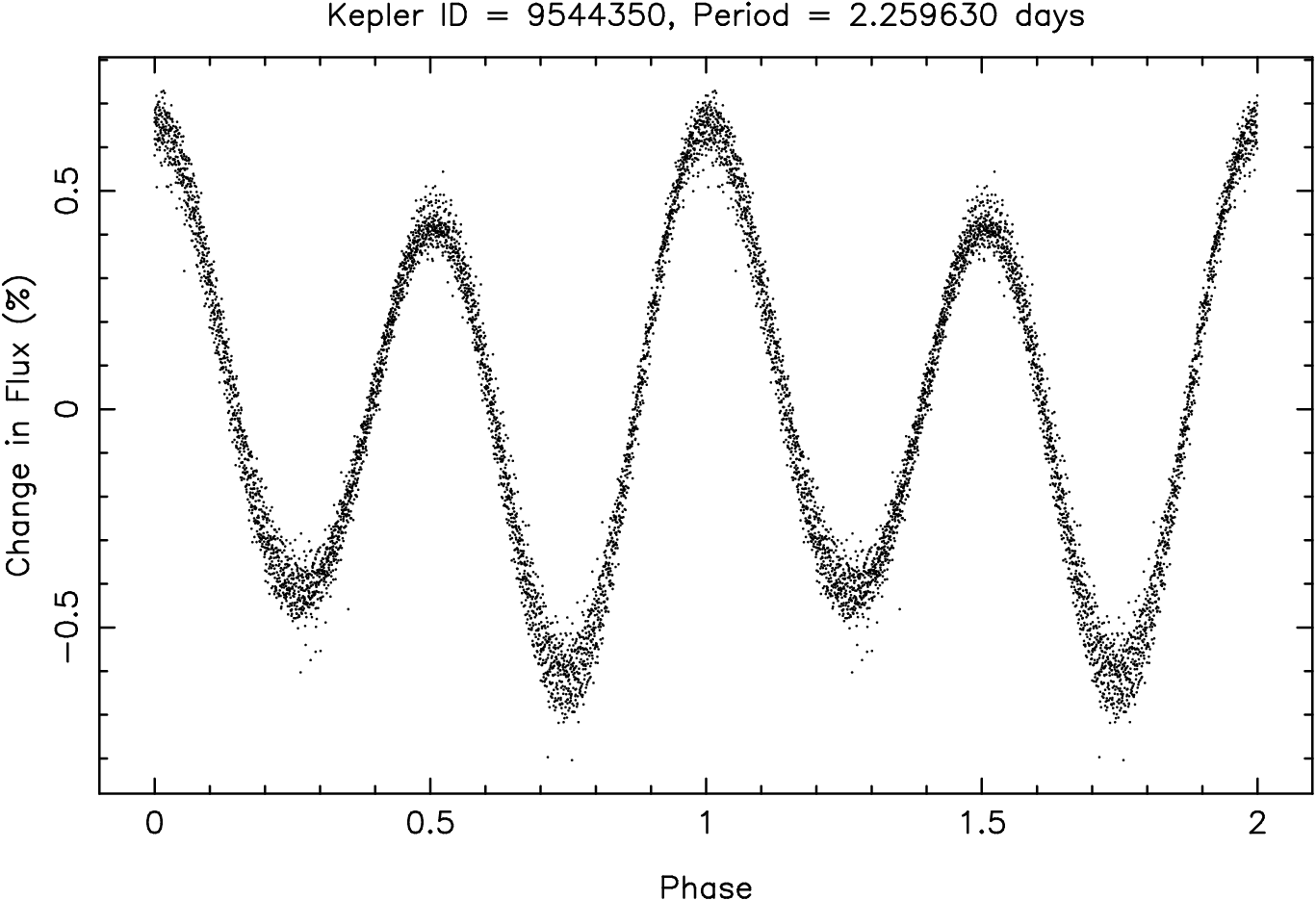}
    \end{tabular}
  \end{center}
  \caption{Light curves from Quarter 2 Kepler data for two cases of
    ellipsoidal variations, Kepler IDs 8386198 (left panel) and
    9544350 (right panel). The fluxes have been converted to
    percentage variations against the mean flux of the data.}
  \label{keplerstars}
\end{figure*}

At the level of photometric precision required here, it is important
to consider the level of intrinsic stellar variability. Depending upon
the properties of the companion, there may be intrinsic stellar
activity which reveals its nature through precision photometry. There
are now hundreds of known brown dwarfs with an increasing
understanding of their atmospheres and activity
\citep{kir11}. However, activity and rotation rates amongst brown
dwarfs and low-mass stars can vary greatly, mostly due to the large
range of temperatures and ages of those objects monitored. Surveys of
relatively young low-mass stars tend to show signs of high
chromospheric magnetic activity and increased rotation rates with
decreasing temperature \citep{bec11,jen09,sch11}. A spectral analysis
of an M dwarf sample by \citet{bro10} showed that increased rotation
appears to be more common in stars later than M3, indicating that
measurable rotational braking is reduced for fully convective
stars. \citet{ber11} investigated the variability of the exoplanet
hosting star GJ~1214 (0.16~$M_\odot$) and find a periodicity of
53~days with an amplitude of 3.5~millimags at the MEarth
bandpass. Clearly if one has the precision to detect planetary phase
variations then one also has the capability to detect stellar activity
from the companion. Indeed it is expected that a low-mass star instead
of planetary companion will show larger chromospheric and coronal
activity compared with equivalent single stars \citep{zaq02}. If the
orbital parameters of the companion are sufficiently understood from
radial velocity measurements then one can extract the variability
signals of the companion and the host star. There may be cases where
the orbital period of the companion is close to the rotation period of
the star, which is generally in the range 10--40~days for radial
velocity host stars \citep{sim10}. In such cases, the peaks in the
power spectrum from a fourier analysis of the photometry may separate
to a degree where starspot variability due to rotation can be
isolated.

This discussion is referring to the activity of a potential
stellar-mass companion. However, the intrinsic stellar variability
could well substantially stronger than these effects. An analysis of
Kepler data by \citet{cia11} found that most dwarf stars are stable
down to the the precision of the Kepler spacecraft, with G-dwarfs
being the most stable of the studied spectral types. The main cause of
photometric variability in F--G--K stars is starspots and rotation, as
verified by the Kepler variability study performed by
\citet{bas11}. Since the orbital period of the radial velocity target
stars is well determined, this will aid in separating the signals of
planetary phase from that of the host star variability whose period is
likely related to the stellar rotation period. It should be noted that
disentangling the variability of the host star may result in an
increased the observing time requirement.

%%%%%%%%%%%%%%%%%%%%%%%%%%%%%%%%%%%%%%%%%%%%%%%%%%%%%%%%%%%%%%%%%%%%

\subsection{Doppler Boosting}

Doppler boosting occurs due to a relativistic effect which creates an
apparent increase and decrease in the light from the host star when it
is moving towards and away from the observer. The fractional amplitude
of the effect is given by
\begin{equation}
  \frac{\Delta F}{F_0} = \frac{(3-\alpha)K}{c}
  \label{boost}
\end{equation}
where $K$ is the radial velocity semi-amplitude and $\alpha$ is the
derivative of the bolometric flux with respect to the frequency in the
stationary frame of reference. This effect has been discussed in terms
of planetary companions and high-precision photometry by \citet{loe03}
and \citet{fai11}. Doppler boosting due to stellar-mass companions has
been detected in the photometry from the Kepler mission
\citep{fai12,van10}. The photometric variability, including Doppler
boosting, induced by KOI-13.01 led to its confirmation as a planet
\citep{maz12,shp11}. It was additionally shown by \citet{loe03} and
\citet{shp11} that the effect of Doppler boosting can be significantly
larger than those from phase and ellipsoidal variations for orbital
periods greater than $\sim 10$~days.

We are considering planets which have been detected with precision
radial velocities. The amplitude of Doppler boosting is directly
proportional to $K$ and thus we remove this effect with only one free
parameter. Note that observing this effect does not remove the $\sin
i$ ambiguity of the companion mass. This is discussed further in
Section \ref{signatures}.

%%%%%%%%%%%%%%%%%%%%%%%%%%%%%%%%%%%%%%%%%%%%%%%%%%%%%%%%%%%%%%%%%%%%

\subsection{The O'Connell Effect}

The O'Connell effect refers to the height difference that may occur
between the maxima in the light curves of close binary stars. There
are several possible causes for this effect, such as starspots or gas
streaming between the binary components. This effect has been
previously studied for eclipsing binary stars by \citet{dav84} and
more recently by \citet{wil09}. Binaries with low-mass stellar
companions are known to exhibit this effect \citep{aus07}, making this
a noteworthy phenomenon here. In particular, the precision of data
from the Kepler mission allows an unprecedented investigation of the
frequency and source of this effect.

In Figure \ref{keplerstars} we show two example light curves which
show several of the effects discussed in this section, Kepler IDs
8386198 and 9544350. These data are corrected flux values from the
Quarter 2 release of the Kepler mission, extracted using the Kepler
interface of the NASA Exoplanet
Archive\footnote{http://exoplanetarchive.ipac.caltech.edu/}. These
stars were included in the Kepler Eclipsing Binary
Catalog\footnote{http://astro4.ast.villanova.edu/aprsa/kepler/} by
\citet{prs11} as having signatures due to ellipsoidal variations but
they can also be seen to have clear O'Connell effect signatures. The
amplitude of these variations are substantial: 5-10 millimags, and so
are well within detection thresholds for stellar-mass companions (see
Figure \ref{ellip}). This is also significantly larger than the
previously mentioned variations and so care must be taken to
distinguish this effect for those more subtle in amplitude.

%%%%%%%%%%%%%%%%%%%%%%%%%%%%%%%%%%%%%%%%%%%%%%%%%%%%%%%%%%%%%%%%%%%%

\section{Observable Signatures}
\label{signatures}

\begin{figure*}
  \begin{center}
    \includegraphics[angle=270,width=14.0cm]{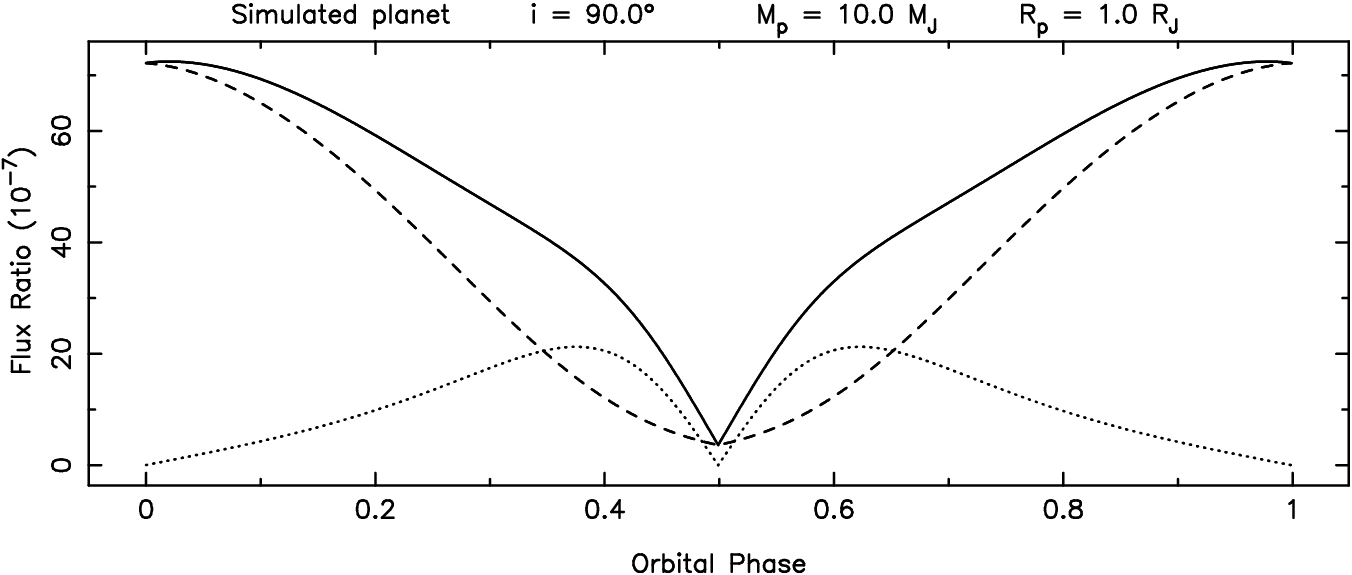} \\
    \includegraphics[angle=270,width=14.0cm]{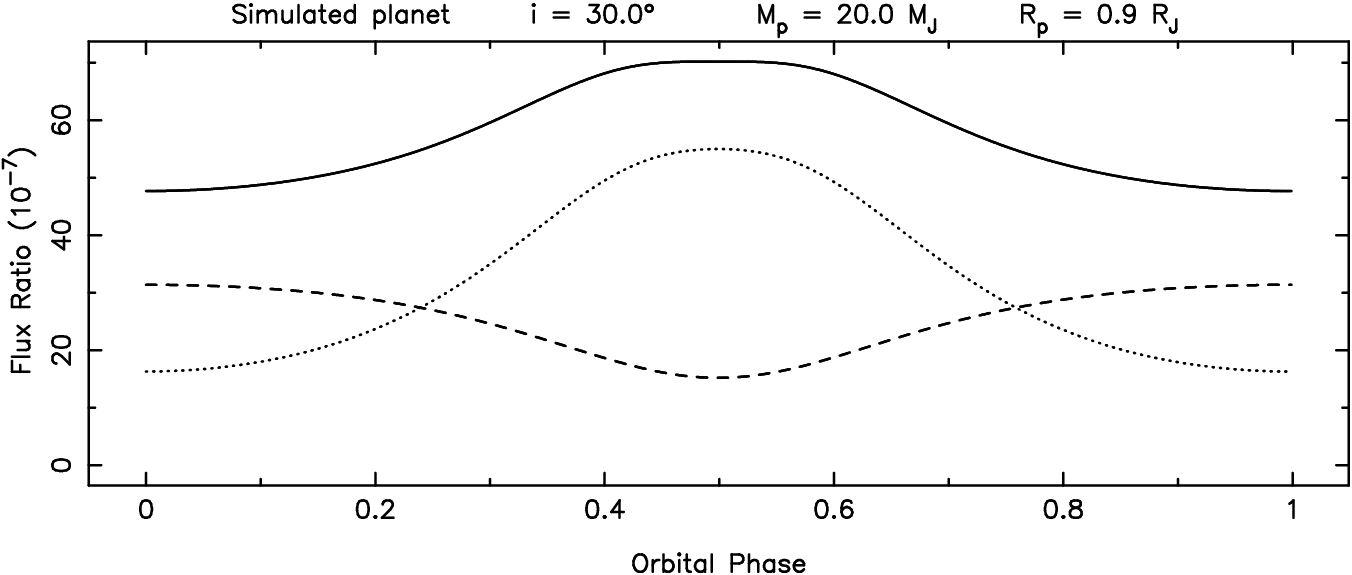} \\
    \includegraphics[angle=270,width=14.0cm]{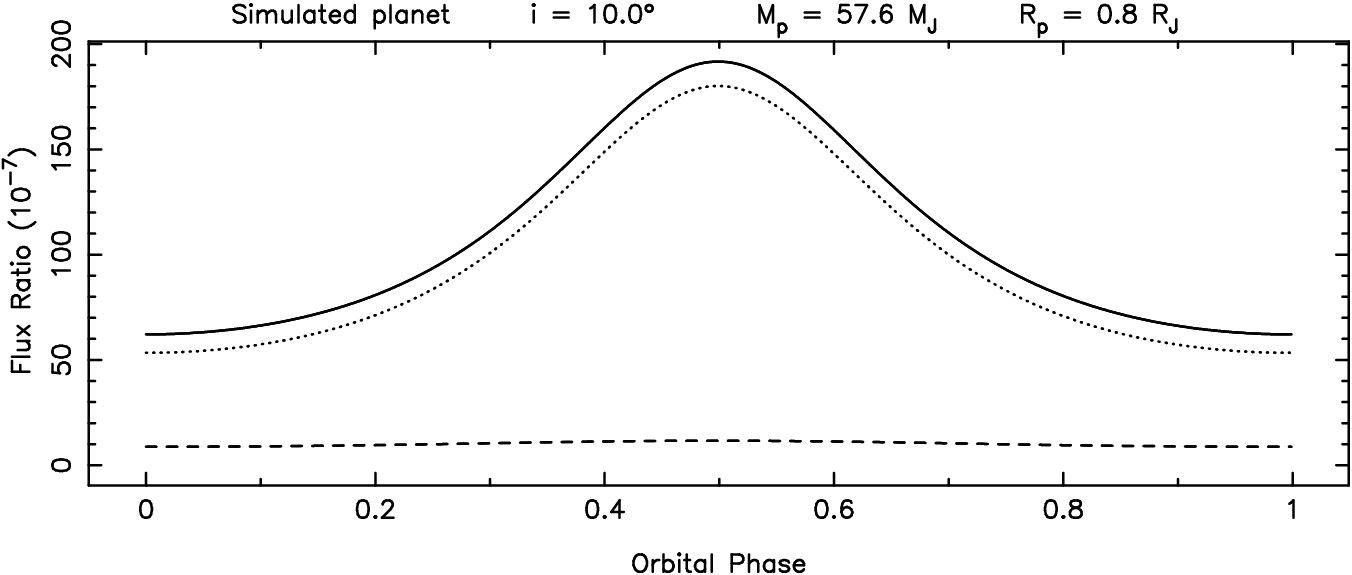}
  \end{center}
  \caption{Photometric signature for a simulated companion in a 5 day
    eccentric ($e = 0.2$, $\omega = 90\degr$) orbit around a
    solar-type star. An orbital phase of zero corresponds to when the
    companion is at superior conjunction. The dashed line is the phase
    variation, the dotted line is the ellipsoidal variation, and the
    solid line is the total combined signature. The three panels show
    the effect of changing the inclination of the system from edge-on
    ($i = 90\degr$, top panel) to the smaller inclinations shown in
    the bottom two panels.}
  \label{sim}
\end{figure*}

Here we combine the effects we have considered in previous sections
and apply them to the two examples discussed earlier, HD~114762b and
HD~162020b. Recall that these various observable signatures are being
considered for companions which are known through their precision
radial velocity detections. In the case of the two examples being
considered here, the radial velocity amplitudes are 612~m\,s$^{-1}$
and 1813~m\,s$^{-1}$ which leads to fractional Doppler boosting
amplitudes of $8.8 \times 10^{-6}$ and $2.6 \times 10^{-5}$
respectively. For the orbits of these planets, we assume that Doppler
boosting effects may be well characterized and we inject the two
remaining and comparable effects of ellipsoidal and phase variations.

\begin{figure*}
  \begin{center}
    \includegraphics[angle=270,width=14.0cm]{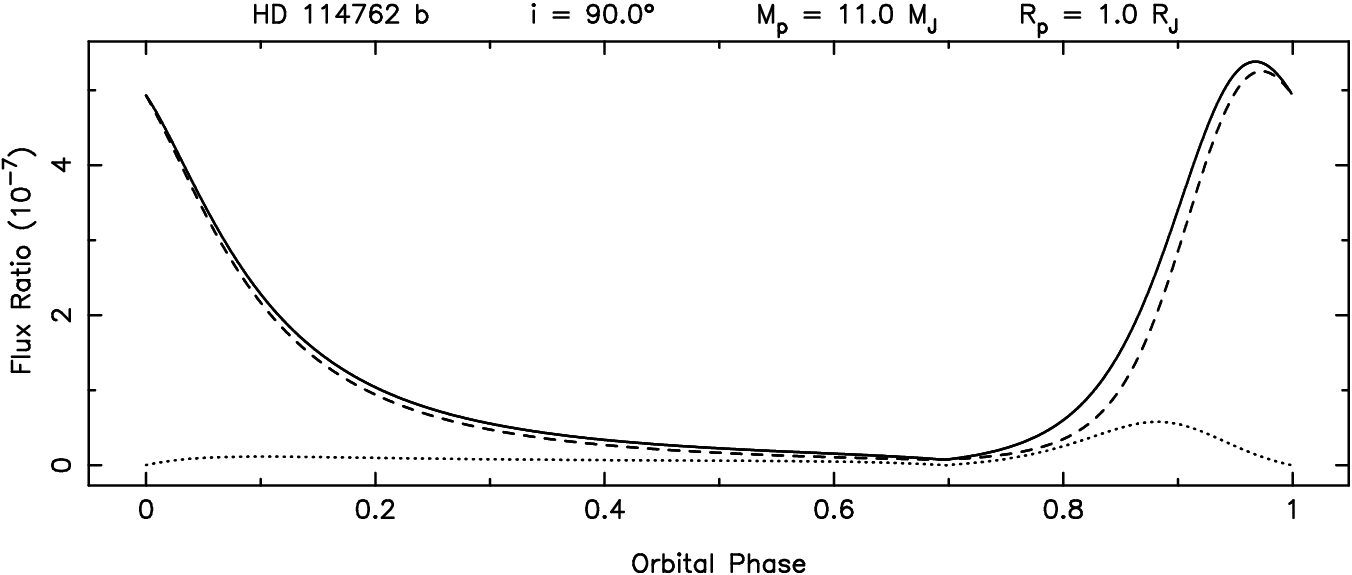} \\
    \includegraphics[angle=270,width=14.0cm]{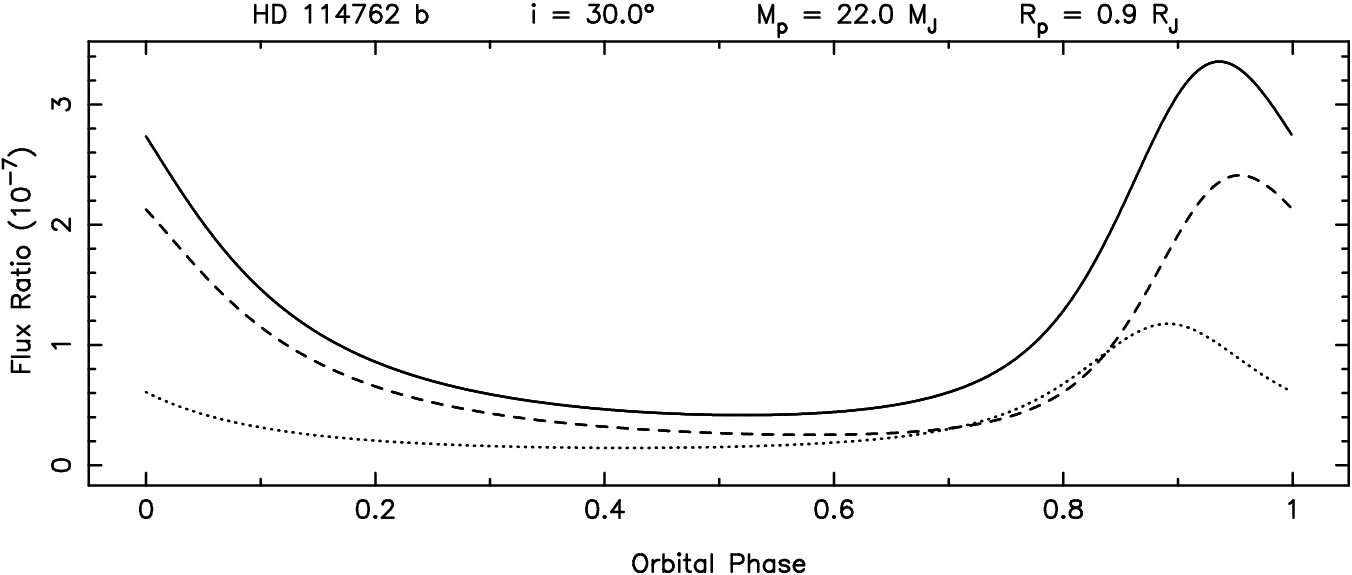} \\
    \includegraphics[angle=270,width=14.0cm]{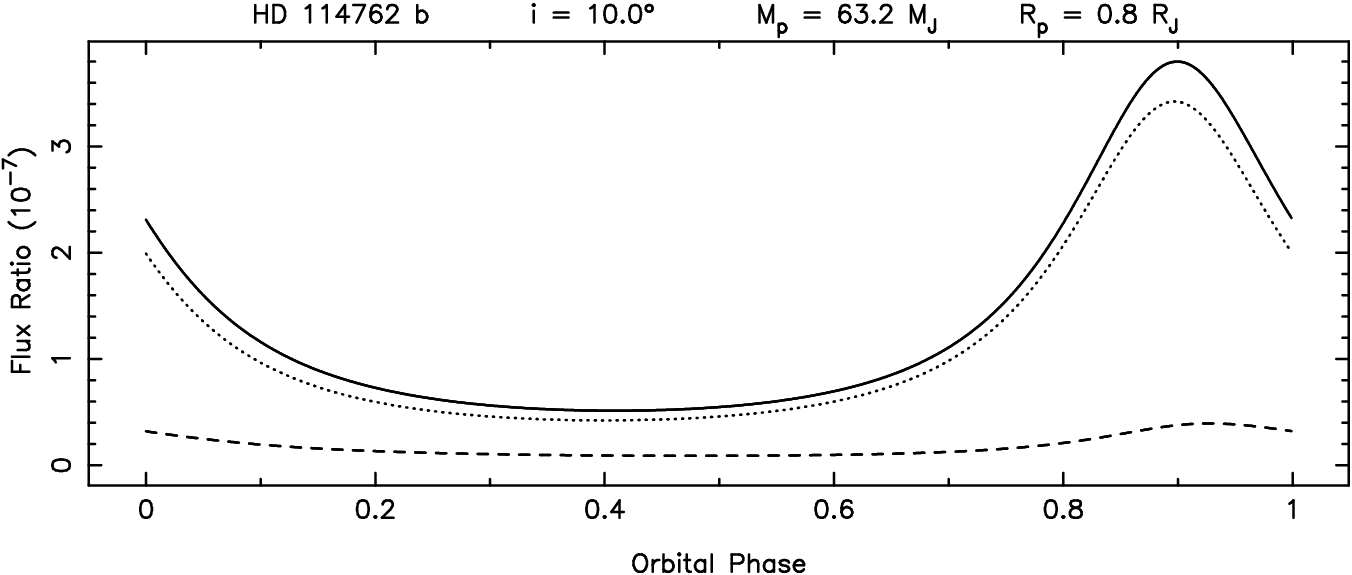}
  \end{center}
  \caption{Predicted photometric signatures for the companion to
    HD~114762. The dashed line is the phase variation, the dotted line
    is the ellipsoidal variation, and the solid line is the total
    combined signature. The three panels show the effect of changing
    the inclination of the system from edge-on ($i = 90\degr$, top
    panel) to the smaller inclinations shown in the bottom two
    panels.}
  \label{hd114762}
\end{figure*}

It was shown by \citet{kan11a} that the phase signature one expects
from a planet in an eccentric orbit is intricately related to both the
inclination and periastron argument of the orbit. If one also includes
the companion mass, radius, and various other photometric signatures
as free parameters then it can become difficult to disentangle the
physical model to account for the observations. Shown in Figure
\ref{sim} is an example sequence of photometric signatures from a
simulated companion in a 5 day orbit around a solar-type star. The
companion has a minimum mass of $M_p \sin i = 10 \ M_J$ and is in an
eccentric orbit ($e = 0.2$) where periastron passage occurs between
the observer and the star ($\omega = 90\degr$). In each panel, the
dashed line is the phase variation, the dotted line is the ellipsoidal
variation, and the solid line is the total combined signature. The
flux ratio refers to the apparent change in flux of the host star,
either by reflected light by the companion or the ellipsoidal
distortion of the stellar shape. An orbital phase of zero corresponds
to when the companion is at superior conjunction. Thus the phase
variation tends to peak near phase zero whereas (for edge-on orbits,
depicted in the top panel) the ellipsoidal variation has two minima:
when the companion is either behind or in front of the star. At this
inclination the companion is considered to be of planetary mass and
the phase signature dominated the photometric variability. The middle
panel shows the effect of reducing the inclination to $30\degr$ such
that the mass of the companion has now doubled. The phase amplitude is
now much reduced and the ellipsoidal distortion of the star as the
companion passes through periastron passage is more visible. At an
inclination of $10\degr$ (shown in the bottom panel) the orbit is now
observed almost face-on. The companion is a high-mass brown dwarf and
the phase signature is almost eliminated, partly due to the slightly
reduced size but mostly due to the reduced contrast between the day
and night-side of the companion as it becomes more self-luminous. Note
also the greatly increased amplitude of the ellipsoidal variations.

%%%%%%%%%%%%%%%%%%%%%%%%%%%%%%%%%%%%%%%%%%%%%%%%%%%%%%%%%%%%%%%%%%%%

\subsection{HD~114762b}

\begin{figure*}
  \begin{center}
    \includegraphics[angle=270,width=14.0cm]{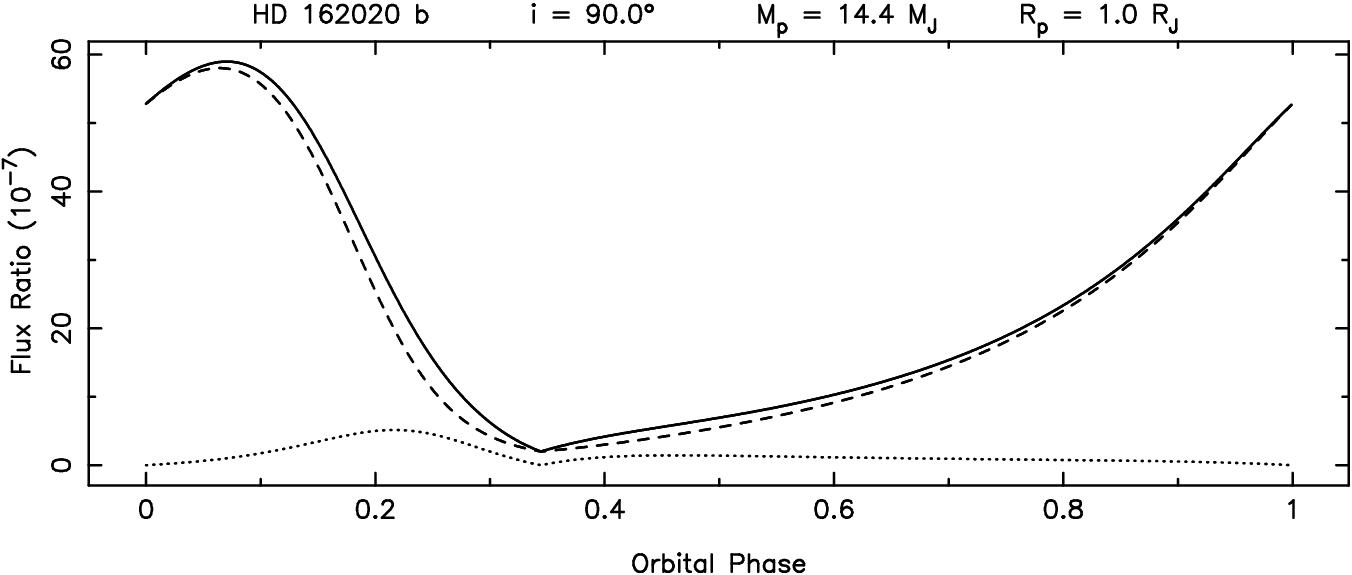} \\
    \includegraphics[angle=270,width=14.0cm]{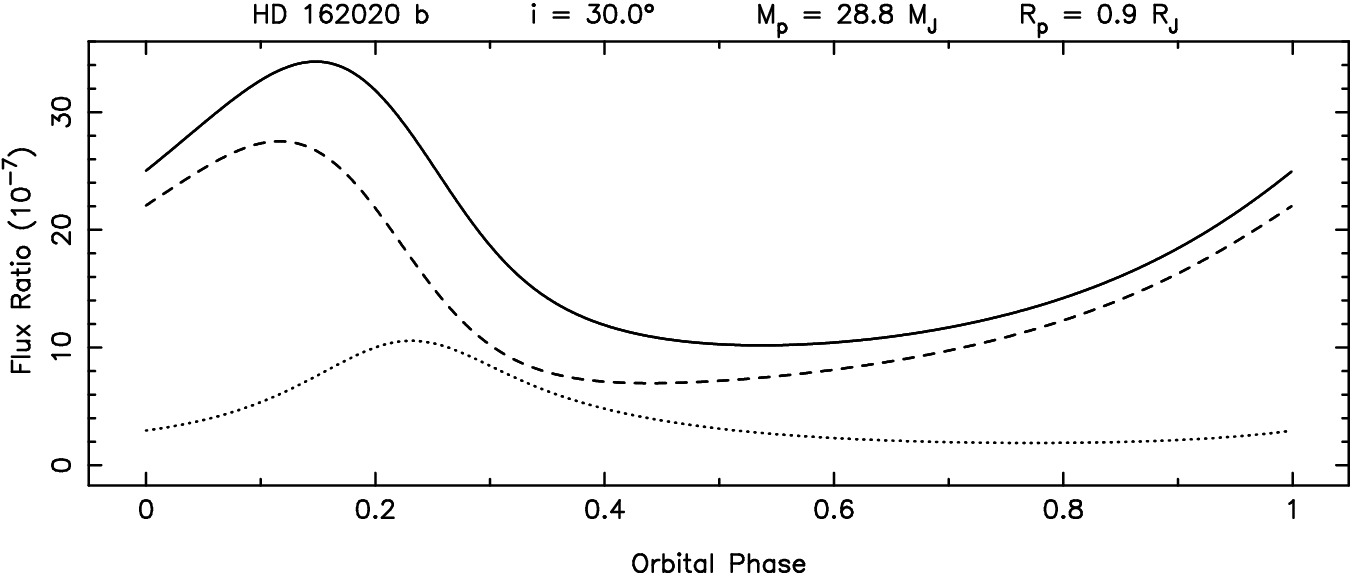} \\
    \includegraphics[angle=270,width=14.0cm]{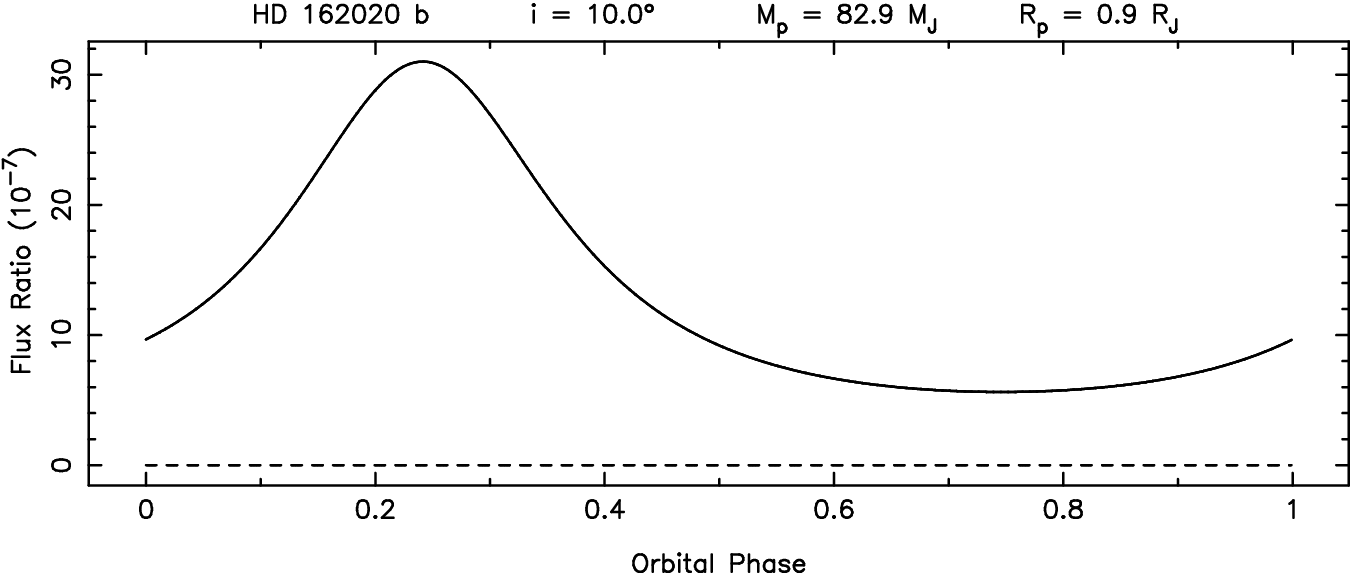}
  \end{center}
  \caption{Predicted photometric signatures for the companion to
    HD~162020. The dashed line is the phase variation, the dotted line
    is the ellipsoidal variation, and the solid line is the total
    combined signature. The three panels show the effect of changing
    the inclination of the system from edge-on ($i = 90\degr$, top
    panel) to the smaller inclinations shown in the bottom two
    panels.}
  \label{hd162020}
\end{figure*}

Shown in Figure \ref{hd114762} are the predicted photometric
variations expected for the companion orbiting HD~114762. We use the
stellar and orbital parameters measured by \citet{kan11c}. This
companion has a minimum mass of $M_p \sin i = 10.98 \pm 0.09 \ M_J$
and is in a $\sim 84$~day period orbit with an eccentricity of
0.34. The periastron argument of $\omega = 201.28\degr$ means that
periastron passage takes place almost behind the star where the
orbital phase is $\sim 0.95$. Recall from Equations \ref{fluxratio}
and \ref{ellipecc} that phase and ellipsoidal variations are
proportional to $r^{-2}$ and $r^{-3}$ respectively. Thus in this case
the phase variations dominate the signature until the mass is pushed
high into the brown dwarf regime as shown in the bottom panel.

There are several aspects of note here. Due to the relatively large
star--planet separation there is only a small change in the total
flux. However, as the phase signature disappears and the ellipsoidal
signature grows the total flux stays approximately the same. Notice
that there is a phase offset between the peak of ellipsoidal and phase
variations. The peak of the ellipsoidal variation depends on the
star--planet separation and the orientation of the major axis of the
distorted stellar profile, whereas the phase variation also depends
upon the phase function of the planet. This is the key which unlocks
the difference between the planetary and stellar companion signatures
in this case since one can detect the change in phase far more easily
than one can detect the change in amplitude of the signature.

%%%%%%%%%%%%%%%%%%%%%%%%%%%%%%%%%%%%%%%%%%%%%%%%%%%%%%%%%%%%%%%%%%%%

\subsection{HD~162020b}

Figure \ref{hd162020} shows the equivalent predicted photometric
variations for the companion to HD~162020. The minimum mass is $M_p
\sin i = 14.4 \ M_J$ and the orbital period and eccentricity are
8.43~days and 0.277 respectively with a periastron argument of $\omega
= 28.4\degr$. These orbital parameters are from the measurements of
\citet{udr02}, who estimate the host star as being K3V. We derive a
radius for the host star by applying the surface gravity $\log g$ from
\citet{val05} to the relation
\begin{equation}
  \log g = \log \left( \frac{M_\star}{M_\odot} \right) - 2 \log
  \left( \frac{R_\star}{R_\odot} \right) + \log g_\odot
\end{equation}
where $\log g_\odot = 4.4374$ \citep{sma05}. From this we calculate a
stellar radius of $R_\star = 0.52 \ R_\odot$. We additionally
calculate the radius using the relations of \citet{tor10} and find
$R_\star = 0.53 \pm 0.03 \ R_\odot$. The ellipsoidal variations are
sensitive to the radius of the host star (see Equation \ref{ellipecc})
but the consistency of the radius determinations and small
uncertainties means that the calculated amplitude is quite robust.

The panels of the figure show that the phase variations are able to
dominate the total signature even when the inclination is decreased to
$30\degr$. However, as the mass rapidly increases beyond this point
the phase variations drop to zero. By the time an inclination of
$10\degr$ is reached it is clear that the companion is able to sustain
hydrogen burning and the photometric variations and now due to tidal
distortions induced by the eccentric orbit. As was the case with
HD~114762b, the total amplitude of the variations does not
significantly change but the phase offset grows as the ellipsoidal
variations emerges as the dominant effect.

%%%%%%%%%%%%%%%%%%%%%%%%%%%%%%%%%%%%%%%%%%%%%%%%%%%%%%%%%%%%%%%%%%%%

\section{A Note on Spectral Line Detection}

An additional effect of increasing the companion mass is the
introduction of absorption bands which are characteristic of brown
dwarfs and late-type stars. One may then wonder if such features would
become apparent in high-precision spectral data or if the companion
and its star could be considered a spectroscopic binary. The optical
spectrum of an M dwarf is typically dominated by TiO and VO absorption
bands. Considerable progress has been made on characterizing the
properties of the cooler L dwarfs. An example of an L dwarf was
detected using both radial velocities and high-resolution imaging
orbiting the solar analog HR 7672 by \citet{liu02}. The L dwarf
prototype, GD~165B, has been used to model the spectra of many similar
types of brown dwarfs through the identification of prominent spectral
features \citep{kir99}. In particular, L dwarfs exhibit strong metal
hydride bands and alkali metal lines. \citet{kir99} found the Na
doublet at 8183, 8195 \AA \ to be especially strong and a possible
candidate for detection in optical passbands. For the even cooler T
dwarfs, such as Gl 229B, the methane absorption bands and broad
absorption features due to alkali metals tend to dominate
near-infrared (NIR) spectra \citep{sau00, sen00}. The major hinderance
facing detection prospects for these features is that they primarily
manifest at red wavelengths where the flux is dramatically reduced at
optical passbands. Thus, resolution of the companion properties
through spectral line detection is better suited to the near-IR
instruments which are being developed to perform searches for
exoplanets around late-type stars.

%%%%%%%%%%%%%%%%%%%%%%%%%%%%%%%%%%%%%%%%%%%%%%%%%%%%%%%%%%%%%%%%%%%%

\section{The Unknown Radius and Albedo}
\label{unknown}

A major source of uncertainty in the discussion thus far results from
the radius and albedo of the planet. These are generally unknown for
non-transiting planets and so we are using models to describe their
dependence on the other measured properties, such as mass. Equation
\ref{fluxratio} shows that the planet-to-star flux ratio depends
linearly on the albedo and quadratically on the planetary
radius. Conversely, these planetary properties have no impact on
either the ellipsoidal or Doppler Boosting variations. The
simulations shown in Section \ref{signatures} demonstrate that the
contribution of the planetary phase to the total variations become
significantly less when the orbit is close to face-on, even for
eccentric orbits. Thus the uncertainties in the radius and albedo
become particularly important for one to be able to discern the
planetary properties for low inclination orbits.

Regarding planetary albedo, there has been considerable effort to
produce analytic models for estimating the albedo at a given
star--planet separation, such as the work of \citet{cah10},
\citet{mad12}, and \citet{sel11}. The situation is more complicated
than a simple distance-albedo relation however, as shown by the recent
discovery of a surprisingly high albedo for Kepler-7b \citep{dem11}.
\citet{cow12} provides a summary of the recent geometric albedo
measurements, the diversity of which may partially be a function of
the heat redistribution efficiency in the high equilibrium temperature
regime. \citet{dem11} attribute the anomalously high albedo of
Kepler-7b as being due to a combination of Rayleigh scattering and
clouds. It is also possible that a smooth dependency of albedo on
star--planet separation is broken by phase transitions whereby removed
reflective condensates reappear in the upper atmosphere as clouds for
a small range of equilibrium temperatures, thus increasing the albedo.
Clearly we require a greater understanding of giant planet albedos in
order to be able to unambiguously extract the phase variations
component for non-transiting planets.

Regarding planetary radius, our understanding of the mass-radius
relationship is developing rapidly as pointed out in Section
\ref{intro}. Based upon the current knowledge of transiting planets,
\citet{kan12} show that the radii of planets with masses greater than
$\sim 0.3$ Jupiter masses follow an approximately linear model. This
mass range also encompasses the majority of non-transiting planets for
which their planetary status is ambiguous. It is expected that the
variation seen within this region is due in no small part to the fact
that highly-irradiated giant planets dominate this sample due to the
bias of the transit technique. However, Kepler is improving our
understanding of planetary radii at a larger range of period than that
which is encompassed by the ground-based surveys. The proposed
Transiting Exoplanet Survey Satellite (TESS) will add to this
understanding but will also benefit from the follow-up potential of
the planet discoveries due to the brightness of the host stars
\citep{ric10}.

Ultimately the results here rely upon the not unreasonable expectation
that our knowledge of planetary radii and atmospheres will improve
dramatically in future years. The combination of this improved
knowledge and the availability of precision photometry will aid
greatly towards breaking the degeneracy in these models.

%%%%%%%%%%%%%%%%%%%%%%%%%%%%%%%%%%%%%%%%%%%%%%%%%%%%%%%%%%%%%%%%%%%%

\section{Feasibility Discussion}

The models presented here show that changing the mass of the companion
will result in unique signatures which can be used to constrain the
mass and subsequent properties of the companion. Such a detection of
these signatures present a significant challenge to instrumentation
requirements and our current understanding of the mass-radius
relationship. The precision requirement for successful detection of
the signatures for the two examples provided is photometry with an
accuracy of $\sim 10^{-7}$. The photometer for the Kepler mission is
designed to achieve high-precision photometry over the 6.5 hour window
of a transit, but is not designed for long-term stability over the
lifetime of the mission \citep{bor10}. The orbits of the radial
velocity planets are well understood in most cases and so we can
accurately predict both the amplitude of the predicted phase signature
but also the phase and times of maximum and minimum flux ratios. In
contrast, the vast majority of Kepler targets are too faint for the
acquisition of accurate radial velocity measurements and so predicting
these times for eccentric orbits amongst Kepler targets is more
difficult. The afore-mentioned TESS mission will provide an
opportunity to perform high precision photometry on these bright host
stars, many of which will have known radial velocity planets. The
expectation that TESS will have sufficient photometric precision to
detect down to Earth-mass planets around F--G--K stars means that this
will be very close to the precision requirements for this
experiment. The planned Lyot coronagraph on NIRCam for the James Webb
Space Telescope (JWST), may be able to achieve phase detections for a
sample of the most favorable targets, though in this case the
instrument is optimized towards young planets around late-type stars.

Ground-based observations face more challenges in terms of correcting
for atmospheric stability. However, future generation telescopes will
provide opportunities to achieve very high precision, such as the
European Extremely Large Telescope (E-ELT), the Thirty Meter Telescope
(TMT), and the Giant Magellan Telescope (GMT). One example of how high
precision can be achieved from such telescopes has been provided by
\citet{col10} whose use of tunable narrow-band filters with the Gran
Telescopio Canarias enabled a photometric precision of $< 0.05$\%.
Further developments of custom filters, adaptive optics systems, and
next-generation telescopes will hopefully provide a competitive
ground-based source for achieving high photometric precision.

A possible alternative approach to detecting the inclination is
astrometric follow-up of these targets, such as that carried out using
Hipparcos data by \citet{ref11}.  The amplitude of the astrometric
signal is roughly proportional to the mass of the secondary component.
We refer the reader to \cite{kan11d} for a more concise comparison of
exoplanet detection methods in the long-period regime.  Thus the
examples shown in Section \ref{massradius} will lead to a factor of
$\sim 5$ increase in the expected astrometric signal. This still
results in astrometric signals at the $\mu$arcsec level and a
challenge to detect from the ground. However, the space-based Gaia
mission is predicted to have a single-measurement astrometric
precision of 5--5.5~$\mu$arcsec \citep{cas08} and will therefore
contribute greatly to sifting high-mass companions from the current
exoplanet sample \citep{soz01}.

%%%%%%%%%%%%%%%%%%%%%%%%%%%%%%%%%%%%%%%%%%%%%%%%%%%%%%%%%%%%%%%%%%%%

\section{Conclusions}

Perhaps the main ambiguity that is inherent to detections of radial
velocity exoplanets is the inclination of the orbit which produces a
measurement of the minimum mass rather than the mass itself. The
methods described here primarily rely on the physical properties of
the companion to distinguish between various classes of orbits, as
opposed to using the orbital properties as proposed by \citet{bla97}.
The different classes of objects which lie along the mass spectrum
from planets to low-mass stars is becoming better understood in terms
of their atmospheres and photospheric activity. However, the change in
mass alone is enough to produce distinct signatures which can
distinguish planets from higher mass objects.

In practice, detection of these signatures is going to be difficult to
accomplish, even if one can achieve the needed precision. Once the
mass increases to the point where hydrogen burning can be sustained,
the variability effects are numerous leading to a veritable plethora
of possible phase curves for the combined system. Note, for example,
that we have not considered the changing gravity-darkening of the host
star due to a variable star--planet distance. The required precision
is becoming achievable however with current and planned space-based
observing platforms. Note that, even though the data from the Kepler
mission is an excellent example of such exquisite precision, the
mission is searching for transits of objects around relatively faint
stars \citep{bor10}. Hence these targets tend to have small prospects
for a radial velocity orbital solution and no ambiguity with regards
to their inclination. The kind of post-discovery analysis suggested
here is therefore more suitable towards missions which can easily
target the bright stars that comprises the bulk of radial velocity
exoplanet host stars. Missions such as TESS will be able to monitor
such stars as well as conducting the survey for new transiting systems
\citep{deming09}.

%%%%%%%%%%%%%%%%%%%%%%%%%%%%%%%%%%%%%%%%%%%%%%%%%%%%%%%%%%%%%%%%%%%%

\section*{Acknowledgements}

The authors would like to thank Davy Kirkpatrick, John Stauffer, David
Ciardi, and William Welsh for several useful discussions. We would
also like to thank the anonymous referee, whose comments greatly
improved the quality of the paper. This research has made use of the
NASA Exoplanet Database, which is operated by the Jet Propulsion
Laboratory, California Institute of Technology, under contract with
the National Aeronautics and Space Administration.

%%%%%%%%%%%%%%%%%%%%%%%%%%%%%%%%%%%%%%%%%%%%%%%%%%%%%%%%%%%%%%%%%%%%


\begin{thebibliography}{}

\bibitem[\protect\citeauthoryear{Anderson et al.}{2011}]{and11}
  Anderson, D.R., et al., 2011, ApJ, 726, L19
\bibitem[\protect\citeauthoryear{Austin et al.}{2007}]{aus07} Austin,
  S.J., Robertson, J.W., Tycner, C., Campbell, T., Honeycutt, R.K.,
  2007, ApJ, 133, 1934
\bibitem[\protect\citeauthoryear{Baraffe et al.}{2003}]{bar03}
  Baraffe, I., Chabrier, G., Barman, T.S., Allard, F., Hauschildt,
  P.H., 2003, A\&A, 402, 701
\bibitem[\protect\citeauthoryear{Basri et al.}{2011}]{bas11}
  Basri, G., et al., 2011, AJ, 141, 20
\bibitem[\protect\citeauthoryear{Becker et al.}{2011}]{bec11} Becker,
  A.C., Bochanski, J.J., Hawley, S.L., Ivezi\'c, Ž., Kowalski, A.F.,
  Sesar, B., West, A.A., 2011, ApJ, 731, 17
\bibitem[\protect\citeauthoryear{Berta et al.}{2011}]{ber11} Berta,
  Z.K., Charbonneau, D., Bean, J., Irwin, J., Burke, C.J., D\'esert,
  J.-M., Nutzman, P., Falco, E.E., 2011, ApJ, 736, 12
\bibitem[\protect\citeauthoryear{Black}{1997}]{bla97} Black, D.C.,
  1997, ApJ, 490, L171
\bibitem[\protect\citeauthoryear{Borucki et al.}{2010}]{bor10}
  Borucki, W.J., et al., 2010, Science, 327, 977
\bibitem[\protect\citeauthoryear{Borucki et al.}{2011a}]{bor11a}
  Borucki, W.J., et al., 2011, ApJ, 728, 117
\bibitem[\protect\citeauthoryear{Borucki et al.}{2011b}]{bor11b}
  Borucki, W.J., et al., 2011, ApJ, 736, 19
\bibitem[\protect\citeauthoryear{Browning et al.}{2010}]{bro10}
  Browning, M.K., Basri, G., Marcy, G.W., West, A.A., Zhang, J., 2010,
  ApJ, 139, 504
\bibitem[\protect\citeauthoryear{Budaj}{2011}]{bud11} Budaj, J., 2011,
  AJ, 141, 59
\bibitem[\protect\citeauthoryear{Burrows et al.}{1993}]{bur93}
  Burrows, A., Hubbard, W.B., Saumon, D., Lunine, J.I., 1993, ApJ,
  406, 158
\bibitem[\protect\citeauthoryear{Burrows et al.}{2007}]{bur07}
  Burrows, A., Hubeny, I., Budaj, J., Hubbard, W.B., 2007, ApJ, 661,
  502
\bibitem[\protect\citeauthoryear{Cahoy et al.}{2010}]{cah10} Cahoy,
  K.L., Marley, M.S., Fortney, J.J., 2010, ApJ, 724: 189
\bibitem[\protect\citeauthoryear{Casertano et al.}{2008}]{cas08}
  Casertano, S., Lattanzi, M.G., Sozzetti, A., Spagna, A., Jancart,
  S., Morbidelli, R., Pannunzio, R., Pourbaix, D., Queloz, D., 2008,
  A\&A, 482, 699
\bibitem[\protect\citeauthoryear{Ciardi et al.}{2011}]{cia11} Ciardi,
  D.R., von Braun, K., Bryden, G., van Eyken, J., Howell, S.B., Kane,
  S.R., Plavchan, P., Stauffer, J.R., 2011, AJ, 141, 108
\bibitem[\protect\citeauthoryear{Claret}{2000}]{cla00} Claret, A.,
  2000, A\&A, 363, 1081
\bibitem[\protect\citeauthoryear{Col\'on et al.}{2010}]{col10}
  Col\'on, K.D., Ford, E.B., Lee, B., Mahadevan, S., Blake, C.H.,
  2010, MNRAS, 408, 1494
\bibitem[\protect\citeauthoryear{Cowan et al.}{2012}]{cow12} Cowan,
  N.B., Machalek, P., Croll, B., Shekhtman, L.M., Burrows, A., Deming,
  D., Greene, T., Hora, J. L., 2012, ApJ, 747, 82
\bibitem[\protect\citeauthoryear{Davidge \& Milone}{1984}]{dav84}
  Davidge, T.J., Milone, E.F., 1984, ApJS, 55, 571
\bibitem[\protect\citeauthoryear{Deleuil et al.}{2008}]{del08}
  Deleuil, M., et al., 2008, A\&A, 491, 889
\bibitem[\protect\citeauthoryear{Demory et al.}{2009}]{dem09} Demory,
  B,-O., et al., 2009, A\&A, 505, 205
\bibitem[\protect\citeauthoryear{Demory et al.}{2011}]{dem11} Demory,
  B,-O., et al., 2011, ApJ, 735, L12
\bibitem[\protect\citeauthoryear{Deming et al.}{2009}]{deming09}
  Deming, D., et al., 2009, PASP, 121, 952
\bibitem[\protect\citeauthoryear{Drake}{2003}]{dra03} Drake, A.J.,
  2003, ApJ, 589, 1020
\bibitem[\protect\citeauthoryear{Faigler \& Mazeh}{2011}]{fai11}
  Faigler, S., Mazeh, T., 2011, MNRAS, 415, 3921
\bibitem[\protect\citeauthoryear{Faigler et al.}{2012}]{fai12}
  Faigler, S., Mazeh, T., Quinn, S.N., Latham, D.W., Tal-Or, L., 2012,
  ApJ, 746, 185
\bibitem[\protect\citeauthoryear{Fernandez et al.}{2009}]{fer09}
  Fernandez, J.M., et al., 2009, ApJ, 701, 764
\bibitem[\protect\citeauthoryear{Fortney et al.}{2007}]{for07}
  Fortney, J.J., Marley, M.S., Barnes, J.W., 2007, ApJ, 659, 1661
\bibitem[\protect\citeauthoryear{Fortney et al.}{2008}]{for08}
  Fortney, J.J., Lodders, K., Marley, M.S., Freedman, R.S., 2008, ApJ,
  678, 1419
\bibitem[\protect\citeauthoryear{Green et al.}{2003}]{gre03} Green,
  D., Matthews, J., Seager, S., Kuschnig, R., 2003, ApJ, 597, 590
\bibitem[\protect\citeauthoryear{Grether \& Lineweaver}{2006}]{gre06}
  Grether, D., Lineweaver, C.H., 2006, ApJ, 640, 1051
\bibitem[\protect\citeauthoryear{Harrison et al.}{2003}]{har03}
  Harrison, T.E., Howell, S.B., Huber, M.E., Osborne, H.L., Holtzman,
  J.A., Cash, J.L., Gelino, D.M., 2003, ApJ, 125, 2609
\bibitem[\protect\citeauthoryear{Jenkins et al.}{2009}]{jen09}
  Jenkins, J.S., Ramsey, L.W., Jones, H.R.A., Pavlenko, Y., Gallardo,
  J., Barnes, J.R., Pinfield, D.J., 2009, ApJ, 704, 975
\bibitem[\protect\citeauthoryear{Johnson et al.}{2011}]{joh11}
  Johnson, J.A., et al., 2011, ApJ, 730, 79
\bibitem[\protect\citeauthoryear{Kane \& Gelino}{2010}]{kan10} Kane,
  S.R., Gelino, D.M., 2010, ApJ, 724, 818
\bibitem[\protect\citeauthoryear{Kane \& Gelino}{2011a}]{kan11a} Kane,
  S.R., Gelino, D.M., 2011, ApJ, 729, 74
\bibitem[\protect\citeauthoryear{Kane \& Gelino}{2011b}]{kan11b} Kane,
  S.R., Gelino, D.M., 2011, ApJ, 741, 52
\bibitem[\protect\citeauthoryear{Kane et al.}{2011}]{kan11c} Kane,
  S.R., et al., 2011, ApJ, 735, L41
\bibitem[\protect\citeauthoryear{Kane}{2011}]{kan11d} Kane, S.R.,
  2011, Icarus, 214, 327
\bibitem[\protect\citeauthoryear{Kane \& Gelino}{2012}]{kan12} Kane,
  S.R., Gelino, D.M., 2012, PASP, 124, 323
\bibitem[\protect\citeauthoryear{Kirkpatrick et al.}{1999}]{kir99}
  Kirkpatrick, J.D., Allard, F., Bida, T., Zuckerman, B., Becklin,
  E.E., Chabrier, G., Baraffe, I., 1999, ApJ, 519, 834
\bibitem[\protect\citeauthoryear{Kirkpatrick et al.}{2011}]{kir11}
  Kirkpatrick, J.D., et al., 2011, ApJS, 197, 19
\bibitem[\protect\citeauthoryear{Kraus et al.}{2008}]{kra08} Kraus,
  A.L., Ireland, M.J., Martinache, F., Lloyd, J.P., 2008, ApJ, 679,
  762
\bibitem[\protect\citeauthoryear{Kraus et al.}{2011}]{kra11} Kraus,
  A.L., Tucker, R.A., Thompson, M.I., Craine, E.R., Hillenbrand, L.A.,
  2011, ApJ, 728, 48
\bibitem[\protect\citeauthoryear{Latham et al.}{1989}]{lat89} Latham,
  D.W., Mazeh, T., Stefanik, R.P., Mayor, M., Burki, G., 1989, Nature,
  339, 38
\bibitem[\protect\citeauthoryear{Liu et al.}{2002}]{liu02} Liu, M.C.,
  Fischer, D.A., Graham, J.R., Lloyd, J.P., Marcy, G.W., Butler, R.P.,
  2002, ApJ, 571, 519
\bibitem[\protect\citeauthoryear{Loeb \& Gaudi}{2003}]{loe03} Loeb,
  A., Gaudi, B.S., 2003, ApJ, 588, L117
\bibitem[\protect\citeauthoryear{Lovis et al.}{2011}]{lov11} Lovis,
  C., et al., 2011, A\&A, 528, 112
\bibitem[\protect\citeauthoryear{Lucy}{1967}]{luc67} Lucy, L.B., 1967,
  ZA, 65, 89
\bibitem[\protect\citeauthoryear{Madhusudhan \& Burrows}{2012}]{mad12}
  Madhusudhan, N., Burrows, A., 2012, ApJ, 747, 25
\bibitem[\protect\citeauthoryear{Mazeh et al.}{2012}]{maz12} Mazeh,
  T., Nachmani, G., Sokol, G., Faigler, S., Zucker, S., 2012, A\&A,
  541, 56
\bibitem[\protect\citeauthoryear{Morris \& Naftilan}{1993}]{mor93}
  Morris, S.L., Naftilan, S.A., 1993, ApJ, 419, 344
\bibitem[\protect\citeauthoryear{Pfahl et al.}{2008}]{pfa08} Pfahl,
  E., Arras, P., Paxton, B., 2008, ApJ, 679, 783
\bibitem[\protect\citeauthoryear{Prsa et al.}{2011}]{prs11} Prsa, A.,
  et al., 2011, AJ, 141, 83
\bibitem[\protect\citeauthoryear{Reffert \& Quirrenbach}{2011}]{ref11}
  Reffert, S., Quirrenbach, A., 2011, A\&A, 527, 140
\bibitem[\protect\citeauthoryear{Ribas}{2006}]{rib06} Ribas, I., 2006,
  Ap\&SS, 304, 89
\bibitem[\protect\citeauthoryear{Ricker et al.}{2010}]{ric10} Ricker
  G.R., et al., 2010, AAS Meeting 215, Bulletin of the American
  Astronomical Society, 42, 459
\bibitem[\protect\citeauthoryear{Saffe et al.}{2005}]{saf05} Saffe,
  C., G\'omez, M., Chavero, C., 2005, A\&A, 443, 609
\bibitem[\protect\citeauthoryear{Saumon et al.}{2000}]{sau00} Saumon,
  D., Geballe, T.R., Leggett, S.K., Marley, M.S., Freedman, R.S.,
  Lodders, K., Fegley, B., Sengupta, S.K., 2000, ApJ, 541, 374
\bibitem[\protect\citeauthoryear{Scholz et al.}{2011}]{sch11} Scholz,
  A., Irwin, J., Bouvier, J., Sip\H{o}cz, B.M., Hodgkin, S.,
  Eisl\"offel, J., 2011, MNRAS, 413, 2595
\bibitem[\protect\citeauthoryear{Seager et al.}{2000}]{sea00} Seager,
  S., Whitney, B.A., Sasselov, D.D., 2000, ApJ, 540, 504
\bibitem[\protect\citeauthoryear{Seager et al.}{2007}]{sea07} Seager,
  S., Kuchner, M., Hier-Majumder, C.A., Militzer, B., 2007, ApJ, 669,
  1279
\bibitem[\protect\citeauthoryear{Selsis et al.}{2011}]{sel11} Selsis,
  F., Wordsworth, R.D., Forget, F., 2011, A\&A, 532, 1
\bibitem[\protect\citeauthoryear{Sengupta \& Krishan}{2000}]{sen00}
  Sengupta, S., Krishan, V., 2000, A\&A, 358, L33
\bibitem[\protect\citeauthoryear{Showman et al.}{2009}]{sho09}
  Showman, A.P., Fortney, J.J., Lian, Y., Marley, M.S., Freedman,
  R.S., Knutson, H.A., Charbonneau, D., 2009, ApJ, 699, 564
\bibitem[\protect\citeauthoryear{Shporer et al.}{2011}]{shp11}
  Shporer, A., et al., 2011, AJ, 142, 195
\bibitem[\protect\citeauthoryear{Simpson et al.}{2010}]{sim10} Simpson,
  E.K., Baliunas, S.L., Henry, G.W., Watson, C.A., 2010, MNRAS, 408,
  1666
\bibitem[\protect\citeauthoryear{Smalley}{2005}]{sma05} Smalley,
  B., 2005, Mem. Soc. Astron. Ital. Suppl., 8, 130
\bibitem[\protect\citeauthoryear{Soszy\'nski et al.}{2004}]{sos04}
  Soszy\'nski, I., et al., 2004, Acta Astron., 54, 347
\bibitem[\protect\citeauthoryear{Sozzetti et al.}{2001}]{soz01}
  Sozzetti, A., Casertano, S., Lattanzi, M.G., Spagna, A., 2001, A\&A,
  373, L21
\bibitem[\protect\citeauthoryear{Spiegel et al.}{2011}]{spi11}
  Spiegel, D.S., Burrows, A., Milsom, J.A., 2011, ApJ, 727, 57
\bibitem[\protect\citeauthoryear{Sudarsky et al.}{2005}]{sud05}
  Sudarsky, D., Burrows, A., Hubeny, I., Li, A., 2005, ApJ, 627, 520
\bibitem[\protect\citeauthoryear{Takeda et al.}{2007}]{tak07} Takeda,
  G., Ford, E.B., Sills, A., Rasio, F.A., Fischer, D.A., Valenti,
  J.A., 2007, ApJS, 168, 297
\bibitem[\protect\citeauthoryear{Torres et al.}{2010}]{tor10} Torres,
  G., Andersen, J., Gim\'enez, A., 2010, A\&ARv, 18, 67
\bibitem[\protect\citeauthoryear{Udry et al.}{2002}]{udr02} Udry, S.,
  Mayor, M., Naef, D., Pepe, F., Queloz, D., Santos, N.C., Burnet, M.,
  2002, A\&A, 390, 267
\bibitem[\protect\citeauthoryear{Valenti \& Fischer}{2005}]{val05}
  Valenti, J.A., Fischer, D.A., 2005, ApJS, 159, 141
\bibitem[\protect\citeauthoryear{van Kerkwijk et al.}{2010}]{van10}
  van Kerkwijk, M.H., Rappaport, S.A., Breton, R.P., Justham, S.,
  Podsiadlowski, P., Han, Z., 2010, ApJ, 715, 51
\bibitem[\protect\citeauthoryear{Watson et al.}{2010}]{wat10} Watson,
  C.A., Littlefair, S.P., Collier Cameron, A., Dhillon, V.S., Simpson,
  E.K., 2010, MNRAS, 408, 1606
\bibitem[\protect\citeauthoryear{Welsh et al.}{2010}]{wel10} Welsh,
  W.F., Orosz, J.A., Seager, S., Fortney, J.J., Jenkins, J., Rowe,
  J.F., Koch, D., Borucki, W.J., 2010, ApJ, 713, L145
\bibitem[\protect\citeauthoryear{Wilsey \& Beaky}{2009}]{wil09}
  Wilsey, N.J., Beaky, M.M., 2009, SASS, 28, 107
\bibitem[\protect\citeauthoryear{Wilson}{1990}]{wil90} Wilson, R.E.,
  1990, ApJ, 356, 613
\bibitem[\protect\citeauthoryear{Winn et al.}{2005}]{win05} Winn,
  J.N., et al., 2005, ApJ, 631, 1215
\bibitem[\protect\citeauthoryear{Zaqarashvili et al.}{2002}]{zaq02}
  Zaqarashvili, T., Javakhishvili, G., Belvedere, G., 2002, ApJ, 579,
  810
\bibitem[\protect\citeauthoryear{Zucker et al.}{2007}]{zuc07} Zucker,
  S., Mazeh, T., Alexander, T., 2007, ApJ, 670, 1326

\end{thebibliography}
\end{document}